# Implementation of a Binary Neural Network on a Passive Array of Magnetic Tunnel Junctions

Jonathan M. Goodwill[1], Nitin Prasad[2,3], Brian D. Hoskins[1], Matthew W. Daniels[1], Advait Madhavan[2,4], Lei Wan[5], Tiffany S. Santos[5], Michael Tran[5], Jordan A. Katine[5], Patrick M. Braganca[5], Mark D. Stiles[1], Jabez J. McClelland[1*]

[1]*Physical Measurement Laboratory, National Institute of Standards and Technology, Gaithersburg, MD, 20899, USA*
[2]*Associate, Physical Measurement Laboratory, National Institute of Standards and Technology, Gaithersburg, MD, 20899, USA*
[3]*Department of Chemistry and Biochemistry, University of Maryland, College Park, MD, USA*
[4]*Institute for Research in Electronics and Applied Physics, University of Maryland, College Park, MD, USA*
[5]*Western Digital Research Center, Western Digital Corporation, San Jose, California, 95119, USA*

The increasing scale of neural networks and their growing application space have produced demand for more energy- and memory-efficient artificial-intelligence-specific hardware. Avenues to mitigate the main issue, the von Neumann bottleneck, include in-memory and near-memory architectures, as well as algorithmic approaches. Here we leverage the low-power and the inherently binary operation of magnetic tunnel junctions (MTJs) to demonstrate neural network hardware inference based on passive arrays of MTJs. In general, transferring a trained network model to hardware for inference is confronted by degradation in performance due to device-to-device variations, write errors, parasitic resistance, and nonidealities in the substrate. To quantify the effect of these hardware realities, we benchmark 300 unique weight matrix solutions of a 2-layer perceptron to classify the Wine dataset for both classification accuracy and write fidelity. Despite device imperfections, we achieve software-equivalent accuracy of up to 95.3 % with proper tuning of network parameters in 15 × 15 MTJ arrays having a range of device sizes. The success of this tuning process shows that new metrics are needed to characterize the performance and quality of networks reproduced in mixed signal hardware.

## I. INTRODUCTION

Over the past decade, artificial intelligence algorithms have achieved human-level performance on increasingly complex tasks at the cost of increased neural network size, computing resources, and energy consumption [1–5]. OpenAI's GPT-3, for example, a state-ot-the-art natural language processor, contains 175 billion parameters and requires $3.14\times10^{23}$ floating point operations to train [6], consuming roughly 190 MWh of electrical energy, roughly the average yearly electrical energy consumption of 16 people in the US [7]. Running these algorithms for inference applications—applications that require the model to make predictions but not learn new information—requires lesser but still overwhelming amounts of energy. This makes them difficult to implement in embedded applications where resources are limited, such as cellphones, self-driving cars, or drones [8–10]. This energy inefficiency is in part due to implementing these algorithms using general-purpose hardware such as central and graphical processing units (CPUs and GPUs).

Because CPUs and GPUs have traditional von Neumann computing architectures, they do not store data in the same spatial location as where computation is carried out. For this reason, energy is consumed in moving the data, and the speed of computation is throttled by the time it takes to shuttle from the storage to the computation location. This so-called von Neumann bottleneck has been shown to be severe on large neural network models, with studies showing the majority of the network time and energy can be expended distributing gradient and model data [11–13].

Algorithmic approaches to lessening the data bottleneck have focused on simplifying neural network models to achieve equivalent accuracy with less memory overhead. Strategies include model compression and sparsification of the synaptic weights [14,15], as well as reducing the precision of weights, with many recent networks performing inference with 4 bits of precision [16–18]. Constantly falling bit precision has fueled interest in taking weight reduction to its logical extreme by using binary neural networks - networks whose synaptic weights can be represented by single bits. Low precision networks have demonstrated similar performance to that of their full-precision counterparts on small datasets, but further improvements are needed to achieve equivalent accuracy on larger datasets [19,20].

The trend towards storing larger models on chip has also driven an increasing effort to develop novel hardware architectures for mitigating the von Neumann bottleneck [21–24]. For example, data access time can be greatly reduced through near- or in-memory computing. Near-memory computing aims to move the data closer to the processing location and use hardware with shorter access times such as static random access memory (SRAM) [25]. Pushing the limits of near-memory architectures, chips have been manufactured with enough onboard SRAM to store more than 40 Gb of data [26], or stacked with through-silicon vias to connect memory and processing chips in 3D [27–29]. Taking this approach to the extreme, in-memory computing carries out calculations directly where the memory resides. Demonstrations of in-memory computing have used dynamic random access memory and SRAM, but less mature emerging non-volatile memories have promise of being lower-power solutions [23,30].

One proposed solution that leverages both low-precision and in-memory computing is to use an array of back-end-of-the-line-compatible magnetic tunnel junctions (MTJs) to implement analog vector matrix multiplication in a binary neural network. Because low-precision computing is more efficient in the analog domain [31], and MTJs are inherently binary and can be designed with low switching energy, they are ideal candidates for minimizing energy consumption in such a hybrid configuration [32,33]. Past investigations have used individual MTJs to experimentally explore the implications of using them in such neural networks [34,35]. A recent demonstration [36] showed a high performance binary neural network using a $64 \times 64$ crossbar array of MTJs integrated with transistors as an active selector device. Such investigations are increasingly showing the utility of using MTJs for computing.

In contrast to active or transistor-integrated arrays, passive, transistorless arrays are potentially an even more efficient way to implement these networks, as they would significantly reduce the additional overhead of transistor capacitances and could be implemented at significantly higher density, while freeing up space for additional transistors that might be needed in peripheral circuitry. Because of the difficulty in fabricating passive nanoscale arrays of MTJs [37,38], favorable performance metrics have only appeared in simulation thus far [39–43]. Here we demonstrate an implementation of a neural network on a passive 15 × 15 crossbar array of MTJs and show the feasibility of obtaining high inference accuracy, even in the presence of hardware imperfections.

Developing a hardware accelerator for inference involves training the neural network offline and transferring the weights to the conductance states of devices. However, because of device non-idealities, it is not possible to exactly reproduce a simulated matrix in hardware and, consequently, it is not possible to know *a priori* what the resultant accuracy of a downloaded network will be. Current methods of increasing the inference accuracy of a downloaded network include optimizing the weights after transfer with further device programming [44,45], optimizing weight mapping onto devices [46], accounting for line resistance voltage drops and parasitics in neural network operations [47,48], and including device variations or noise in the training algorithm to make the final model in hardware more robust to device nonidealities [49,50]. Here, we demonstrate the plausibility of this last approach. We produce many variations (300) of weight matrices using different weight initializations during offline training. In this way, based on array-specific non-idealities, certain weight matrix solutions achieve higher inference accuracy than others. In principle, one would expect networks that better reproduce the target network to achieve higher accuracy.

By programming all 300 weight matrix solutions into the hardware, we are able to quantify the impact of device non-idealities on the distribution of achievable accuracies. We calculate our ability to accurately reproduce each network model through the root mean square (RMS) deviation between the model and the implementation. By optimizing the network conductance-to-weight conversion, we achieve a median accuracy of 95.3 % over all programmed solutions. One finding with implications for embedded inference is that the network parameters that maximized the network's experimental performance are different in general from those that theoretically maximized it. Specifically, the magnitude of the weight normalization constant (see Accuracy Optimization section) that minimized the RMS deviation did not also maximize the accuracy. This result suggests the necessity of new approaches for embedded inference with off-line trained networks on imperfect hardware, and that accurate network recreation is not the ideal criterion for maximizing network performance.

## II. RESULTS

### A. Neural network hardware acceleration with arrays of MTJs

Experiments were carried out on 15 × 15 passive crossbar arrays (no integrated transistors or selection devices) with MTJ diameters of 30 nm, 40 nm, 50 nm, and 60 nm. A scanning electron microscope (SEM) micrograph of the array layout is shown in Fig. 1(a). Details of the MTJ array fabrication can be found in Appendix A. The crossbar arrays are fabricated without integrated control circuitry, so all measurements are made through port-to-port measurements using source measure units and a switch matrix. This approach allows for detailed characterization and control of individual devices, as we describe below, but does not allow us to control and characterize 15 voltages and currents simultaneously.

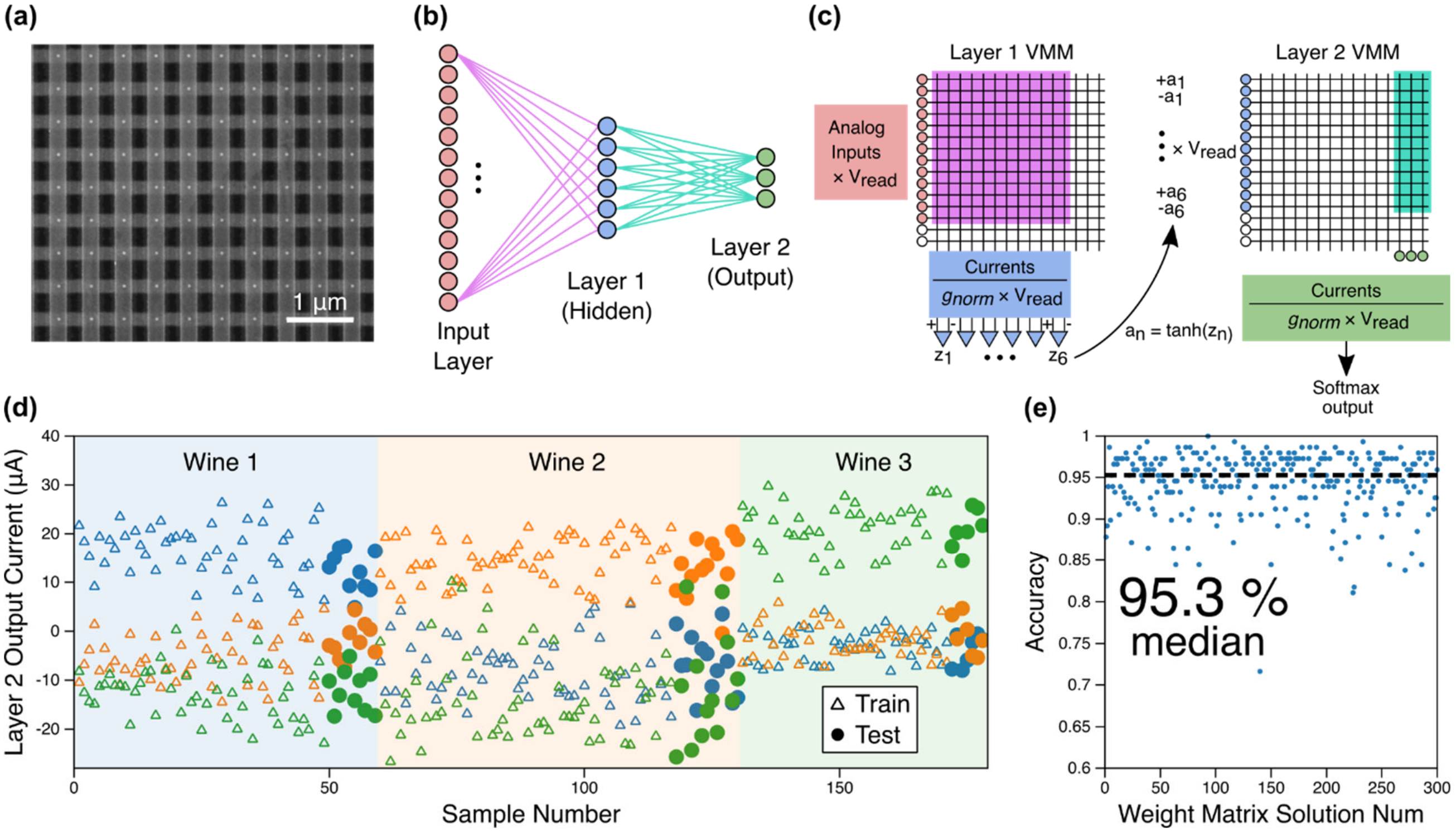

FIG. 1. Classification of the Wine dataset using a 15 × 15 MTJ array. (a) Scanning electron microscope (SEM) micrograph of the MTJ array. (b) Neural network architecture used to classify the Wine dataset samples, containing 13 input neurons, 6 hidden layer neurons (layer 1), and 3 output neurons (layer 2). The hidden layer and output layer neurons each use two MTJs to implement three-level weights. (c) Schematic of the neural network mapping to the MTJ hardware. The hardware equivalent neural network function is the same color as the corresponding function shown in (b). Layer 1 outputs $z_n$ $(n = 1 - 6)$ are transformed by a tanh activation into layer 2 inputs $a_n$. (d) Currents simulated on the output columns of layer 2 in the MTJ array over 148 training and 30 test inputs for a single weight matrix solution with 99.3 % and 93.3 % accuracy on the training and test datasets, respectively, after programming weight values into devices. (e) Optimized classification accuracy on the training dataset of 300 different weight solutions tried in the array. Measurements for (d) and (e) were both performed on an MTJ array with a device diameter of 30 nm.

The dataset used for classification was the Wine dataset [51], which included 178 samples of wine. Each sample has 13 recorded characteristics (for example, alcohol concentration, color intensity, etc.) and an associated label for the cultivar from which the wine was produced.

To avoid trivial convergences of the learned weights towards the class-centroids, a simple 2-layer network was constructed. The architecture of the neural network is shown in Fig. 1(b) and the mapping to hardware shown in Fig. 1(c). The neural network includes 13 input neurons, 6 hidden neurons, and 3 output neurons (one for each of the possible cultivars), producing a 13 × 6 weight matrix for layer 1 and a 6 × 3 weight matrix for layer 2. Consequently, we fit our entire network into the 15 × 15 array, necessarily limiting its size, in contrast to Ref. [36], where the same array was used to 28 times to emulate a large network through reprogramming the same array during the forward pass. Drawing inspiration from the inhibitory and excitatory synapses found in the human brain, we chose to implement weights with two MTJ devices. The weight of the dual MTJ synapse is proportional to the conductivity difference between the two MTJs, thus allowing us to implement negative weights. In layer 1, weights were implemented with adjacent devices arranged left-right, but in layer 2 they were arranged adjacently up-down. This was done to maximize array utilization. However, there is a subtle difference in operation between layers 1 and 2 because of this. More specifically shown in Fig. 1(c), the arrangement of implemented weights in layer 1 requires the difference in columns on the output to be taken, whereas in layer 2 the difference on the output is not necessary. Instead, both positive and negative input values are required on the rows in layer 2. Both methods carry out exactly the same function, just in a different manner. Network training was performed offline, and the learned weights were subsequently downloaded into the MTJ crossbar by serially programming individual devices.

The inference accuracy was determined by reading all device conductance states after programming and using these to scale the currents in software and simulate the number of correctly predicted wine classifications. After writing weights to the crossbar and measuring all effective port-to-port conductances, which include device non-idealities, line resistances, sneak paths, etc., we have all the information necessary to determine the accuracy the solution would produce in an inference process. The inference process itself, involving summing currents to carry out vector matrix multiplications (VMMs), normalizing, and passing through activation functions, can be carried out in software, since we have verified that these steps do not introduce too much noise or uncertainty to the outcome to invalidate the results. We have verified that the applied voltages are in a regime where device conductances have no measurable voltage dependence and our measurements of this network satisfy the superposition principle of linear circuits (see Appendix B). Simulating the full vector-matrix multiply using the individually measured port-to-port device properties reduces the need for the additional external electronics, and focuses most directly on the performance of the passive MTJ crossbar array itself. Additionally, measuring these device properties allows for more thorough analysis of the hyperparameter tuning required to achieve software-equivalent accuracy. This approach is both quantitatively and qualitatively different from the one taken in Ref. [36] for a few reasons. Unlike traditional binary neural networks, as implemented in [36], we opted to only binarize the weights rather than the signals as well. Consequently, we are modeling the transmission and activation of continuously valued signals on a binary network. In addition, we implicitly assume a classic analog to digital conversion of the current through a transimpedance amplifier, though we don't explicitly include this in our model. In Ref. [36], the researchers explicitly implemented a circuit for converting currents into a temporal code by charging a capacitor and counting clock cycles before passing into a software activation function. Such temporal codings are potentially more energy efficient than traditional analog to digital conversion. In principle, such an approach could also be used for our passive array; however, it would require a more careful analysis of the bit precision, and would not be expected to change the results of our analysis provided the device behavior is sufficiently linear.

To simulate the classification of an individual wine sample, the 13 wine attributes (inputs), normalized to be between 0 and 1, were first transformed into voltages by multiplying by a constant voltage $V_{\text{read}}$ (0.2 V); this is the voltage at which each port-to-port device conductance was read in hardware after serial programming. The VMM for layer 1 was then carried out using these voltages on the rows of the array to calculate the currents on the columns. The currents were normalized into dimensionless quantities by dividing by the product of $V_{\text{read}}$ and a conductance hyperparameter $g_{norm}$. The 6 outputs of layer 1 ($z_1 \ldots z_6$) were obtained by taking the difference between adjacent columns and adding a bias. Layer 1 outputs were then fed through a hyperbolic tangent activation function to obtain layer 2 inputs ($a_1 \ldots a_6$). Both positive and negative values of each $a_n$ were multiplied by $V_{\text{read}}$ to use as input voltages on adjacent rows. In layer 2 the output currents were again normalized to dimensionless quantities and a bias was added before being fed through a softmax activation to determine the network classification prediction. For each input, a correct result was tallied whenever the appropriate output current was the largest of the three.

In Fig. 1(d), the simulated current values from each of the three output columns of the MTJ array are shown over all 148 training samples and 30 test samples for a single weight matrix solution. The accuracy is 99.3 % on the training set and 93.3 % on the test set. This high level of performance is obtained after optimizing the $g_{norm}$ hyperparameter, as discussed in the Accuracy Optimization section below. The accuracies on the training dataset for all 300 unique weight matrix solutions trained offline and programmed into the MTJ array are shown in Fig. 1(e). The maximum accuracy is 100 %, the minimum is 71.6 %, and the median is 95.3 % over all solutions. These results demonstrate that a high-accuracy inference binary neural network can be realized using a non-ideal passive MTJ hardware array. Note the classification accuracy of the neural network on the training dataset using the MTJ array is an important metric because it shows how well the network in hardware can represent the network as it was trained in software. All weight matrix solutions trained offline in software achieved a simulated accuracy above 96 % on the training dataset and 95 % on the test dataset, but due to write errors and device non-idealities, perfect software-equivalent accuracy could not be guaranteed after transferring the weights to hardware. Figure 1(e) shows the extent to which the hardware imperfections play a role. As expected, certain solutions performed better than others, but overall the fidelity is sufficient to allow for software-equivalent accuracy on average.

### B. MTJ device and array characterization

An MTJ is formed by stacking two ferromagnetic layers, referred to as the fixed and free layers, together with a thin insulating layer in between. While the magnetization of the fixed layer is pinned, the free layer magnetization can be either parallel or anti-parallel to the fixed layer. The direction of the free layer can be switched by applying a suitable write current through the MTJ, which creates a spin-transfer torque [52]. When the free layer magnetization is parallel (anti-parallel) to that of the fixed layer, the MTJ conductance is high (low). In the subsequent discussion, we refer to the high and low conductance states as the on-state and off-state, respectively, characterized by their conductance values $g_{on}$ and $g_{off}$. The relative conductance is characterized by a tunnel magnetoresistance ratio ($TMR$) defined as

$$TMR \equiv \frac{g_{on} - g_{off}}{g_{off}}. \quad (1)$$

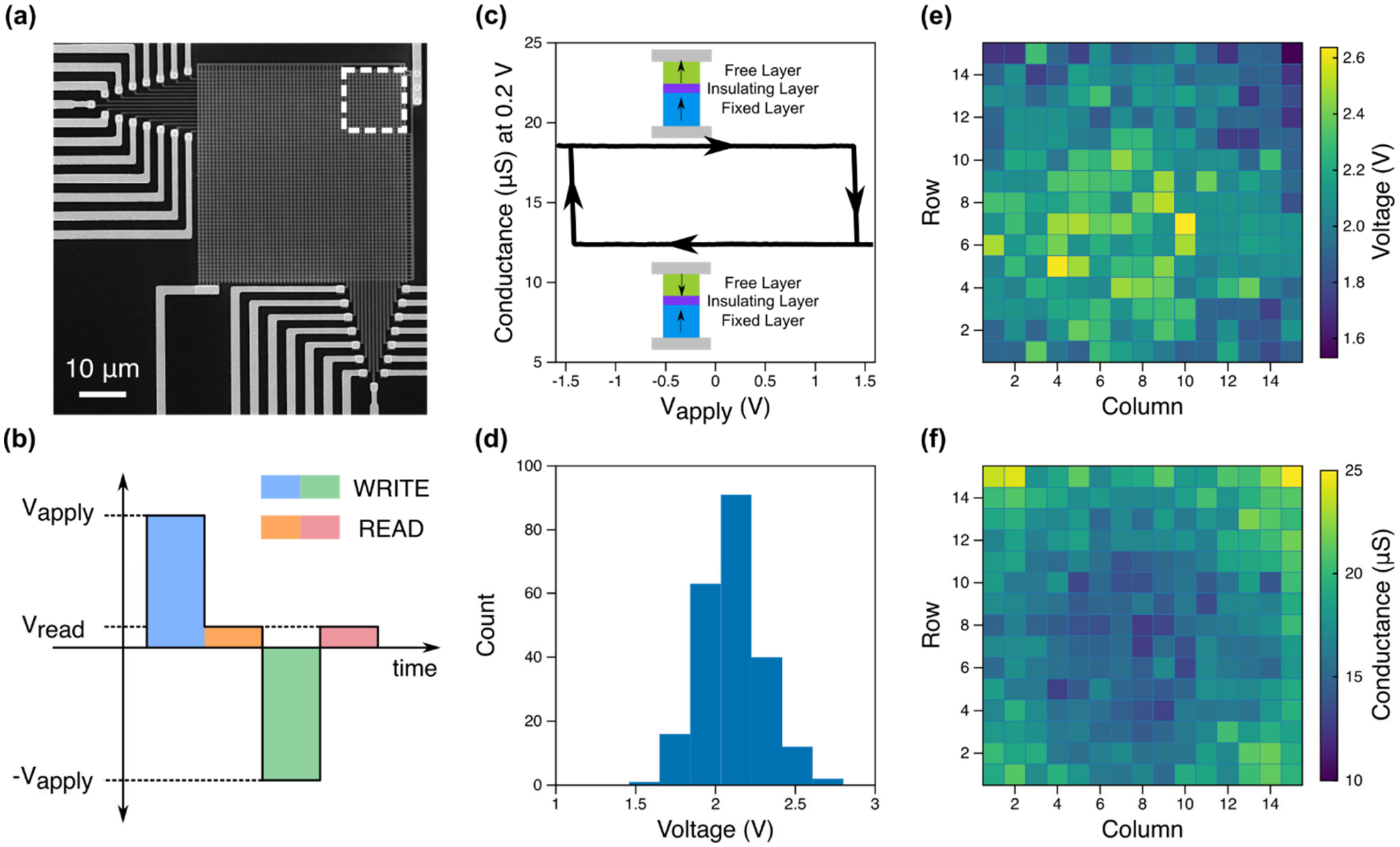

FIG. 2. Neural network hardware. (a) Zoomed out SEM image, showing the 15 × 15 MTJ array (dashed white box) as part of a larger, unused array. Lighter traces are the metal routing lines. (b) Pulse sequence used for writing and verifying devices with alternating write and read biases. The switching voltage for each device is measured by repeating this sequence while increasing $V_{\text{apply}}$ until the device switches. (c) Switching curve of a device (diameter 30 nm) with a 1.5 V threshold from the anti-parallel (off) state to the parallel (on) state. The conductance was measured at a fixed 0.2 V bias. Also shown are the MTJ configurations in each state. (d) Histogram of switching voltages on the 30 nm diameter array, showing essentially a normal distribution. (e) Color plot of the measured switching voltages in the 30 nm diameter array. The voltages are higher near the center where the line resistance is greatest. (f) Color plot of the measured on-state conductances of the MTJs in the array. The values are lower near the center where the line resistance is highest.

The MTJ array is accessed with a probe card that interfaces with an offboard switch matrix and three source-measure units. The metal routing to the array rows and columns is shown in the zoomed out image of the array (Fig. 2(a)) with the active array region indicated by the white dashed box. To write the devices, we use a “V over 2” scheme, which applies $V_{\text{apply}}/2$ to the target column and $-V_{\text{apply}}/2$ to the target row while grounding all other connections. This ensures $V_{\text{apply}}$ is applied to the target device while only half the bias is applied to all the others [53].

Device conductance states were always read by applying a voltage ($V_{read}$) of 0.2 V on the target row with all other connections grounded and measuring the current on the target column.

We use a write-verify scheme, shown in Fig. 2(b), to accurately write device states. This scheme utilizes a sequence of four pulses (1 ms pulse width) where the first and third pulses write the device state with opposite polarities. The first write pulse always attempts to write the device to the opposite of the target state, and the third pulse attempts to switch to the target state. For example, if the target state is the on-state, the first pulse attempts to write to the off-state and the third pulse attempts to write to the on-state. The second and fourth pulses read the device state after each write pulse. A device is ensured to be in the target state by checking the conductance on/off ratio obtained from the second and fourth pulses. This 4-pulse sequence is repeated with increasing $V_{apply}$ until the on/off ratio condition is met or a maximum voltage limit is reached. The maximum applicable voltage is limited to twice the smallest switching voltage in the array to eliminate the risk of switching unwanted devices. If the on/off ratio is still negligible at the maximum voltage limit, the write is deemed unsuccessful and considered a write error for programming the array. The write accuracy was 100 % for the 30 nm array shown here but decreased to 85 % as the device size increased to 60 nm.

The four-pulse write-verify scheme is used instead of a two-pulse scheme because the on/off ratio criteria is more reliable than the device conductance, which could vary between on/off switching cycles. In addition, the four-pulse scheme is more resilient to cycle-to-cycle write errors. For instance, if a device is already in the on-state and the target state is also the on-state, the calculated on/off ratio would be close to unity and lead to needlessly increasing $V_{apply}$. We avoid such a circumstance by always writing the opposite of the target state first so as to validate that the device switched to the correct state on each programming step.

Figure 2(c) shows the conductance read at 0.2 V as a function of increasing $V_{apply}$ for an individual MTJ in the array with device diameter 30 nm. Also shown on the graph are illustrations of the configuration of the free and fixed layers for the corresponding on- and off-states. For this measurement, $V_{apply}$ was swept from -1.6 V to 1.6 V and, as indicated by the vertical transitions, the free layer magnetization direction flipped at roughly ±1.5 V.

At the array level, additional complexities arise because of subtle differences from device to device. Figure 2(d) plots the histogram of effective switching voltage while Figs. 2(e) and 2(f) show the individual device voltage and on-state conductance values for a 15 × 15 MTJ array with device diameter 30 nm as a function of device row and column. The values are "effective" because no device can be separated from the array and tested in isolation; the values for each device are only obtainable from measurements on the device word and bit lines. The data in Figs. 2(d) – 2(f) were obtained with each device originally in the off-state and then subjected to our write-verify scheme to program every device to the off-state again. We measure individual device characteristics when the rest of the array is in the off-state because, as a passive array, the measured properties of each device state are influenced by the states of other devices in the surrounding environment. As more devices are switched to the on-state, it becomes more difficult to measure individual device characteristics due to the increased contribution from sneak paths [54,55].

Of note in these figures are the variations present in both voltage and on/off conductance states. The switching voltage appears to follow a normal distribution with a mean value around 2.2 V, but the map of voltages in Fig. 2(e) indicates that variations do not occur uniformly across the array. Lower voltages and higher conductance values occur towards the periphery of the array, especially at the corners. Similarly, higher switching voltages are required near the center of the array, where devices tend to have lower on-state conductance values. This effect is due to a combination of line resistance and device-device variations. Device-device variations due to minute differences in processing conditions mainly account for the small differences in voltage and conductance between adjacent devices. Line resistance, on the other hand, accounts for systematic differences across the array. Because of the nanoscale size of the metal word and bit lines, the line resistance is non-negligible, and significant voltage drops occur across the lines [48]. This gives the appearance that a device requires higher switching voltage, when in reality it may require a comparable switching voltage, but additional voltage is needed to compensate for the increased drop associated with the line. In the routing configuration of the metal lines in the fabricated arrays, the longest metal lines are on the center row and column, while the shortest lines are on the periphery (see Fig. 2(a)). This is the main effect giving rise to the distributions shown in Figs. 2(d) – 2(f). Similar distributions are observed in the other fabricated sizes of MTJ arrays (see Fig. S1). In general, the switching voltage and standard deviation increased with increasing MTJ diameter, as can be seen in Figs. S2(a) and S2(b). This trend can be anticipated from the increasing importance of the line resistances as the devices' resistances decrease with increasing diameter.

### C. Weight mapping to hardware and inference accuracy of 300 solutions

To encode the weight matrix into the MTJ array, each weight is represented by two adjacent MTJ devices, in a scheme inspired by excitatory and inhibitory synapses. The conductances of these two devices are denoted $g_e$ and $g_i$. The weight is defined as the difference between $g_e$ and $g_i$, divided by a normalization conductance $g_{norm}$. During offline training, weights were given values of {-1, 0, 1} to replicate the possible combinations of $g_e$ and $g_i$ MTJ pair states, as shown by the magnetization orientations in Fig. 3(a).

The weight arrangement for layer 1 utilizes rows 1 - 13 and columns 1 - 12 of the MTJ array with weights arranged as adjacent devices left-right, whereas layer 2 utilizes rows 1 - 12 and columns 13 - 15 with weights arranged as adjacent devices up-down. In all cases the device on the left (layer 1) or top (layer 2) of the pair is excitatory and the device on the right (layer 1) or bottom (layer 2) is inhibitory. This is clarified in Fig. 3(b), which shows the weight mapping of the entire array for both neural network layers, where $g_e$ devices are labeled in orange and $g_i$ devices labeled in blue. The last two rows and last three columns of row 13 are not used in either layer of the neural network, and thus are always written to the off-state.

The dimensionless-equivalent quantity to the current on each column is the sum of weights multiplied by inputs for each neuron. Mathematically, the VMM operation for the $k$th layer manifests as

$$\vec{y}^{k}_{outputs} = \frac{\vec{I}^{k}_{columns}}{V_{read} \cdot g_{norm}} = \left[\vec{x}^{k}_{inputs} \cdot V_{read}\right] \cdot \frac{1}{V_{read} \cdot g_{norm}} \left[\hat{G}^{k}_{e} - \hat{G}^{k}_{i}\right], \quad (2)$$

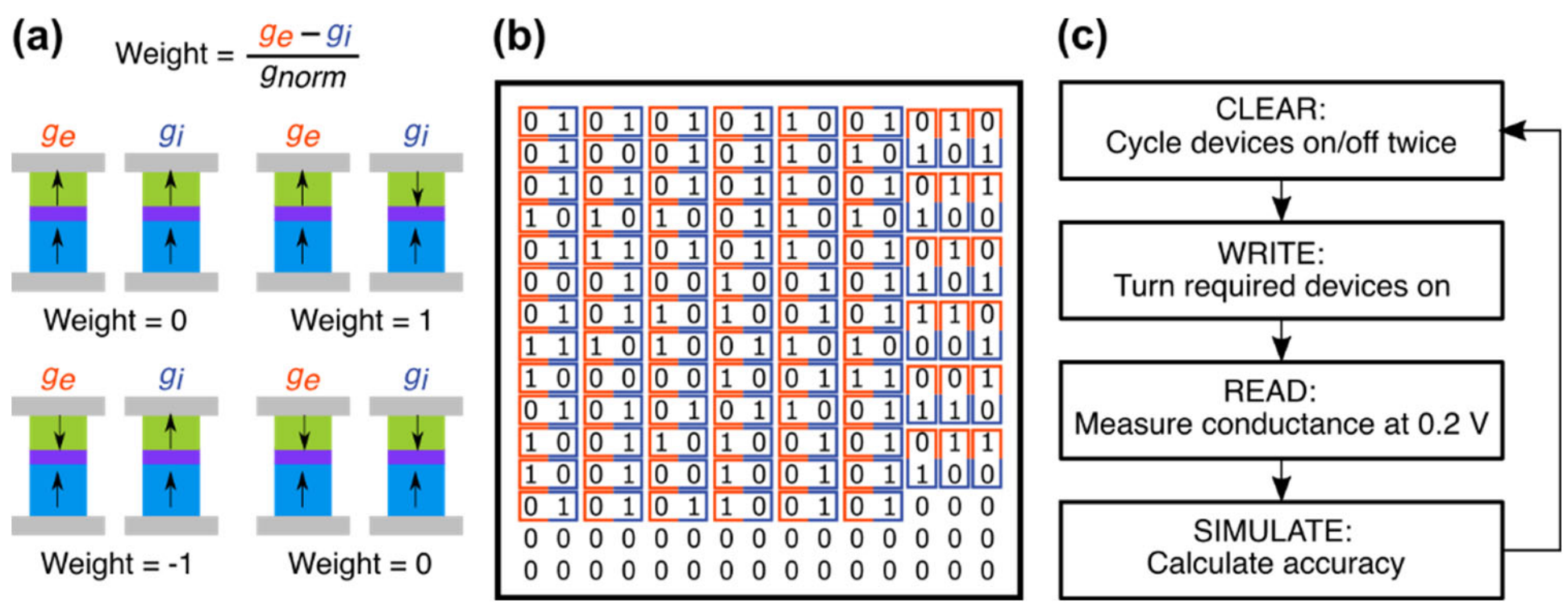

FIG. 3. Writing 300 unique weight solutions and testing inference accuracy. (a) Illustration of how weights were encoded in the hardware using adjacent devices ($g_e$ devices are labeled in orange and $g_i$ devices labeled in blue). (b) Layout of the nearest neighbor differential weight mapping. In the first layer, differential weights are defined by neighboring columns but in the second layer they are defined within neighboring rows. (c) Block diagram of the write algorithm to program the arrays and test inference accuracy.

where $\vec{x}^k$ and $\vec{y}^k$ are dimensionless inputs and outputs, and $\hat{G}^k_{e/i}$ are the respective excitatory and inhibitory weight matrices. The quantity $\frac{1}{g_{norm}}\left[\hat{G}^k_e - \hat{G}^k_i\right]$ is a matrix of the dimensionless weights, which we denote as $\hat{U}^k$ , with dimensionless matrix elements $U^k_{ij}$ that should be directly comparable to the matrix elements of the ideal weight matrix $W^k_{ij}$.

To measure the inference accuracy of all 300 unique weight matrix solutions, each arrangement of weights has to be individually written as conductances into the MTJ array. This was done using the programming approach shown in Fig. 3(c). All operations in Fig. 3(c) utilize the write-verify scheme discussed above and illustrated in Fig. 2(b). Because devices in the on-state decrease write accuracy by increasing sneak-path parasitics, each weight matrix was programmed from an initial state of all devices in the off-state. To ensure this was the case, the first step in writing a particular weight solution was to write all devices to the off-state twice, also called the "clear operation." After clearing the array, only the devices that were required to be in the on-state for that specific weight matrix solution were written during the "write operation." Once written, the "read operation" was carried out by serially reading the conductance of each device at 0.2 V without disturbing the written states. These conductance values were used to calculate the effective weights stored in the MTJ array. Finally, the inference accuracy was determined in the "simulate" operation by calculating the number of correctly predicted wine categories out of the 148 training samples assuming the weight values dictated by the measured conductance states during the "read operation." Results of this procedure are shown in Fig. 4 and were also used to produce Figs. 1(d) and 1(e).

The 15 × 15 maps displayed in Figs. 4(a) and 4(b) demonstrate the high write accuracy achieved by the programming sequence. Figure 4(a) shows the target MTJ device states for a particular weight matrix solution with 1 being the on-state and 0 the off-state. Figure 4(b) shows the corresponding conductance values obtained during the "read operation" on the MTJ array with device diameter 30 nm. As shown previously in Fig. 2(f), there is still variation in the conductance values across the array, but the devices in the on-state can nevertheless be distinguished from devices in the off-state. This distinction is made clear in Fig. 4(c) which shows histograms for the on- and off-states measured during the "write operation." Both on- and off-states have roughly normal distributions with an on/off ratio around 2. The standard deviation of the off-state was smaller than the on-state and the distributions slightly overlap near 14 μS. Similar on/off- conductance state distributions were observed in other MTJ sizes, but importantly, the overlap in on/off-states worsened as MTJ diameter increased, as shown in Figs. S2(c) and S2(d). This had detrimental consequences on the ability to accurately clear and write device states in larger MTJ sizes, as shown in Figs. S3(a) and S3(b).

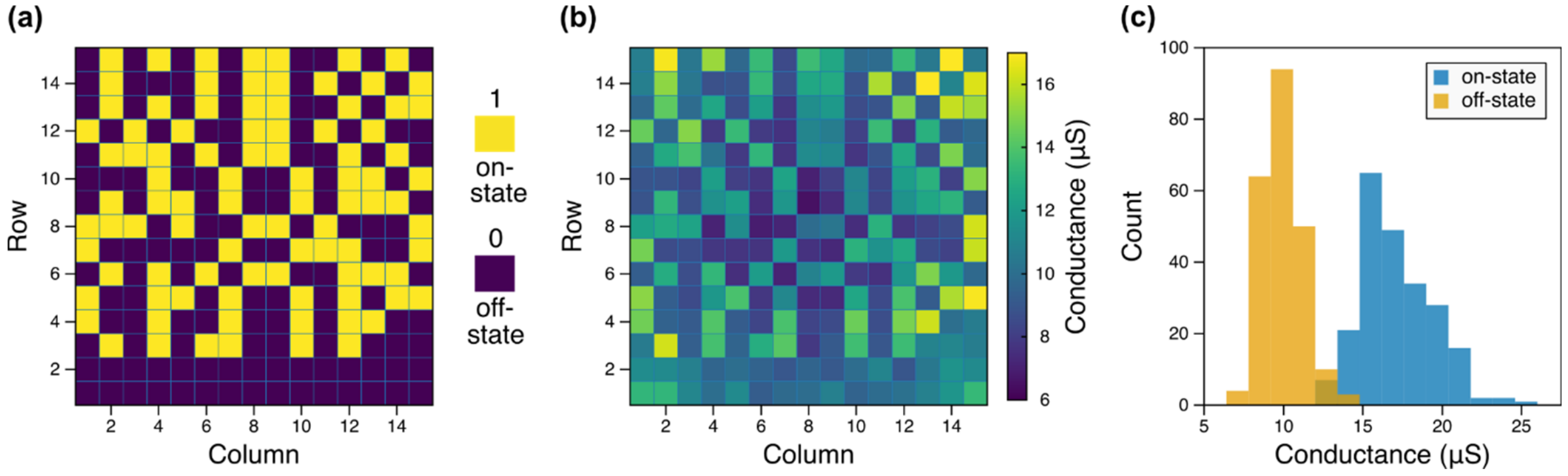

FIG. 4. Writing weight matrices and measuring conductance. (a) A target matrix solution in the passive MTJ array. (b) The "read operation" results of the programmed crossbar to the target array (diameter 30 nm) using the write-verify scheme. (c) Conductance values of each device measured in the on- and off-state during the "write operation." The histogram shows separation of the states.

### D. Accuracy optimization

In this study we used the normalization conductance $g_{norm}$ as a hyperparameter to optimize the classification accuracy. By tuning $g_{norm}$, we could change how well the real weights represented the ideal weights determined in software, and this affected the accuracy distribution over all 300 weight matrix solutions. In the ideal case where all devices have the same $g_{on}$ and $g_{off}$, if all $g_e$, $g_i$ pairs are normalized by the same $g_{norm} = g_{on} - g_{off}$, the weight values reduce to the pure binary values {-1, 0, 1}. This is not the case for a hardware realization because no two devices have the exact same $g_{on}$ and $g_{off}$. Thus, modifying $g_{norm}$ can improve how well some real weights replicate their ideal counterparts, but can also degrade the fidelity of others. This tradeoff means there is an optimal $g_{norm}$ that minimizes the deviation between the real and ideal weights. It also implies that there should be a $g_{norm}$ that maximizes the accuracy. One might expect that these two values of $g_{norm}$ would coincide.

An important note is, just like device non-idealities, the optimal value of $g_{norm}$ for a given hardware implementation is impossible to know *a priori*. In our case, $g_{norm}$ can be optimized because we use measured conductance states of all devices and simulate the accuracy of the neural network at different values of $g_{norm}$. This is made simpler by the fact that the MTJ device conductance does not change as a function of voltage for low applied voltages, making the current on the columns easily computable as the voltage on the input changes over all wine input samples. Because of this dependence, we are also able to test devices individually and add the currents, as if we applied all voltages on all the rows at the same time. It should be noted, however, that this procedure would not work for crossbars with highly nonlinear elements, such as two-terminal selectors. Likewise, at higher biases where the MTJs are more nonlinear, we would expect to see deviations from our calculations.

Simulating the accuracy for any type of hardware neural network becomes more difficult as the number of devices increases and the device conductance changes as a function of voltage. Hyperparameter optimization of the weight mapping to hardware becomes prohibitive due to the large computational requirements of simulation. In most cases the normalization hyperparameter is estimated [56]. For example, a relatively good approximation of $g_{norm}$ can be calculated trivially as:

$$g_{norm} = \bar{g}_{on} - \bar{g}_{off}, \tag{3}$$

where $\bar{g}_{on}$ and $\bar{g}_{off}$ are the average values of $g_{on}$ and $g_{off}$ for the array. This is a much simpler computation to carry out, but as we will show in the following discussion, it turns out to be a poor choice, providing strong incentive for determining the optimal $g_{norm}$.

The accuracy distribution of the 300 unique weight matrix solutions for 30 nm devices is shown as a function of $g_{norm}$ in Fig. 5(a). For these devices, the estimated value of $g_{norm}$ using average $g_{on}$ and $g_{off}$ values was 7 µS. This value of $g_{norm}$ has an inference accuracy with a median of only 60.8 %. Surprisingly, a $g_{norm}$ of 3.4 µS did a much better job of compensating for the array characteristics and resulted in a much higher optimized median accuracy of 95.3 %.

In Fig. 5(b) we show the RMS deviation between the ideal and measured weights as a function of $g_{norm}$ over the 300 weight matrix solutions. The total RMS deviation, $\Delta_{\mathrm{RMS}}$, for a given weight matrix solution of the two neural network layers is calculated as:

$$\Delta_{\mathrm{RMS}} = \sum_{k=1,2} \sqrt{\sum_{ij} \left(W_{ij}^{k} - U_{ij}^{k}\right)^2} \tag{4}$$

where $W_{ij}^{k}$ are the ideal weights for node $ij$ in network layer $k$ and $U_{ij}^{k}$ are the weights for the programmed MTJ array determined from the measured conductances scaled by $g_{norm}$. The superscript $k$ indicates the associated neural network layer and the summation is performed over the weight matrix indices. The point of minimum RMS deviation is significant because this is where the measured weights best reproduce the ideal weights. Figure 5(b) shows that $\Delta_{\mathrm{RMS}}$ is minimized at $g_{norm}$ = 5.5 µS, a value that differs from the accuracy-optimized value and is much closer to the estimated value of 7 μS. The two different optimized values of $g_{norm}$ are highlighted by the vertical dashed lines in Fig. 5(b).

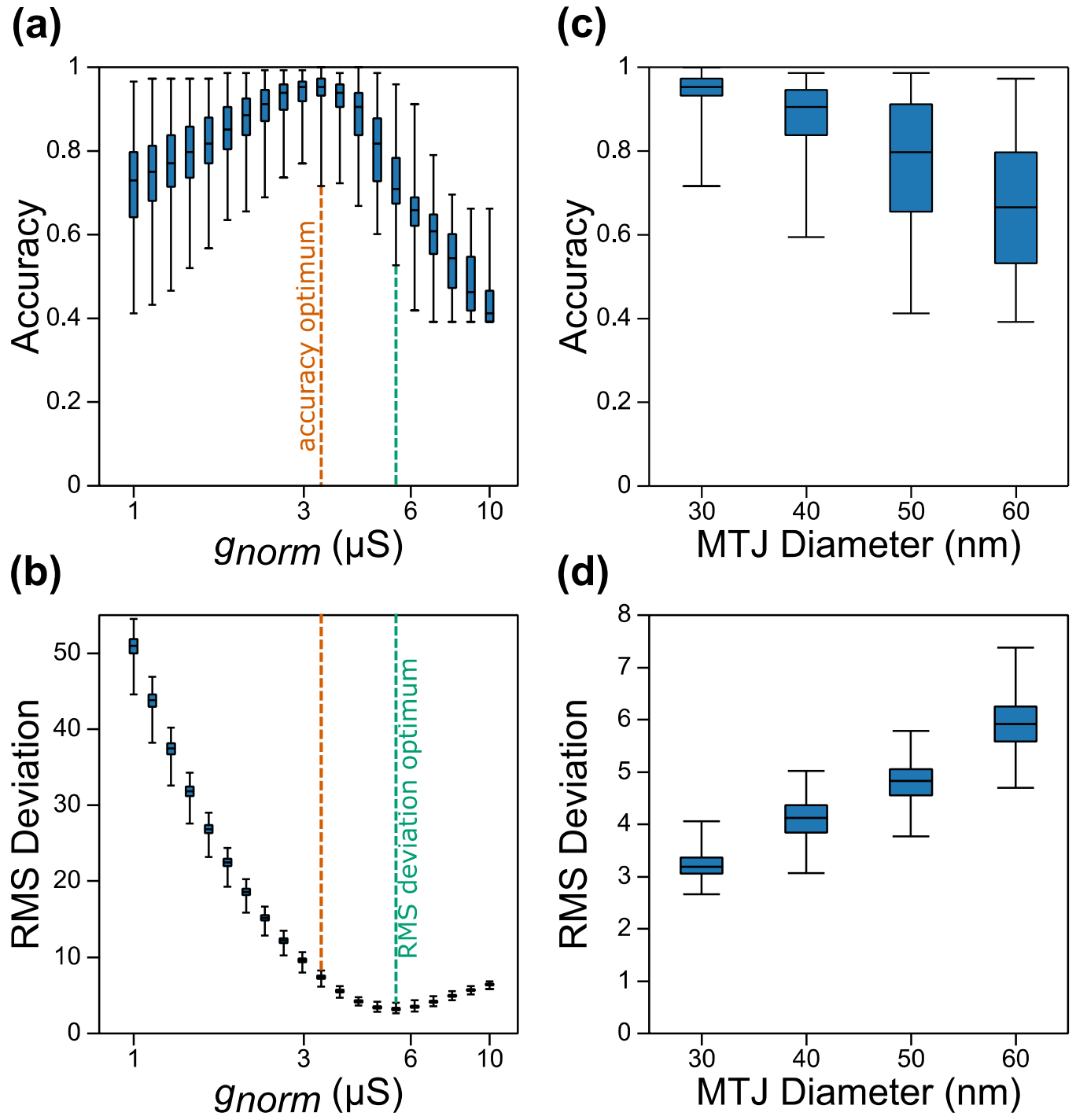

FIG. 5. Optimization of accuracy and RMS deviation over $g_{norm}$ and MTJ diameter. Box-whisker plot of (a) accuracy vs. $g_{norm}$ and (b) RMS deviation vs. $g_{norm}$ for all 300 weight solutions in the 30 nm diameter array. Vertical dashed lines indicate the optimum values for the two different criteria (3.4 μS and 5.5 μS) determined from the median values at each $g_{norm}$. (c) Box-whisker plots of the distributions of optimized accuracies as a function of MTJ diameter. (d) Box-whisker plots of the distributions of minimum RMS deviations as a function of MTJ diameter. In each figure, whiskers indicate maximum and minimum values, whereas box edges represent 25 % and 75 % quartiles and the middle line is the median (50 % quartile).

Figures 5(c) and 5(d) show distributions of the optimized accuracy and minimum RMS deviations over all 300 weight matrix solutions for all four MTJ device sizes fabricated. The full set of plots for accuracy and RMS deviation as a function of $g_{norm}$ are shown in Fig. S4 in the supplemental material. Figure 5(c) shows that the maximal median achievable accuracy decreases as MTJ size increases, while Fig. 5(d) indicates that the minimum RMS deviations simultaneously increase with MTJ size. We attribute these trends to the increasing overlap of on/off conductance states as the device size increases (see Figs. S2(c) and S2(d)), which affects the clear and write accuracy of the array. Although the maximum median accuracy did not occur at the same value of $g_{norm}$ as the minimum median RMS deviation for any device size, the trend of the minimum median RMS deviation does provide an indication of the trend of maximum median accuracies, with overall poorer reproducibility predicting lower accuracy.

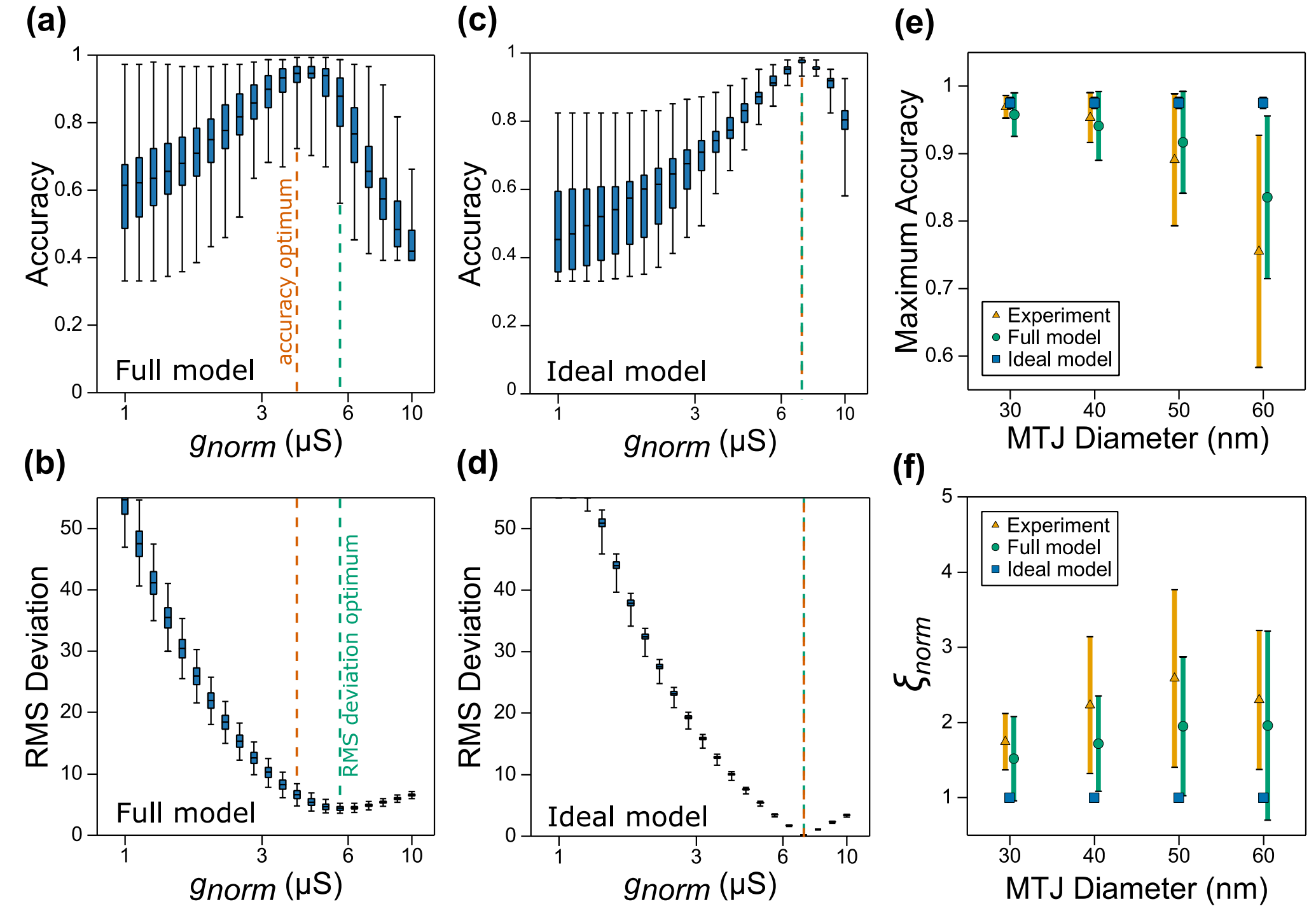

FIG. 6. Accuracy and RMS deviation as a function of the scaling parameter between measured and model networks, $g_{norm}$, and MTJ diameter. Panels (a-d) show the results of simulations with all device variations and line resistances included (full model) and also with no non-idealities (ideal model). Results are for a single realization of device conductance variations of the 30 nm diameter MTJ array. Shown are box-whisker plots of the distribution of results over the 300 unique weight solutions. Panels (a) and (b) show accuracy and RMS deviation, respectively, as a function of $g_{norm}$ for the full model. Panels (c) and (d) show the same for the ideal model. For each box-whisker plot (a-d), whiskers indicate maximum and minimum values, whereas box edges represent 25 % and 75 % quartiles and the middle line is the median (50 % quartile). Panels (e) and (f) show the optimized accuracies and values of $\xi_{norm}$ (see text) for experimental results and the results of simulations like those in (a-d) for different MTJ diameters. The symbols give the mean values. For the experimental results and the ideal model simulation results, the whiskers give the one-standard-deviation width of the distribution of results over the 300 different solution matrices. For the full model results, the whiskers give the one-standard-deviation width of the distribution over 9000 solutions (300 solution matrices for each of 30 realizations of device conductance variations).

To shed light on the reduction of maximum achievable accuracies in experiments and to explain the mismatch between the value of $g_{norm}$ that maximizes the accuracy and the value of $g_{norm}$ that minimizes RMS deviation, we carried out circuit (SPICE) simulations of inference on an MTJ array. The simulations capture the variations in the two-port conductances by accounting for external resistances, line resistances, and random variations in the MTJ properties. The model parameters were obtained by fits to measured data. For each MTJ size, 30 separate realizations of the device variations were implemented by treating the off-state conductance and $TMR$ of each MTJ as independent normally distributed quantities consistent with the measured distributions of values. Figures 6(a) and 6(b) show the full model of the simulated accuracy and RMS deviation as a function of $g_{norm}$ in the presence of line resistances for a representative realization of device-device variations in the 30 nm diameter array. The simulation reproduces the experimental finding that network accuracy is not maximized at the value of $g_{norm}$ that minimizes the RMS deviation. For comparison, Figs. 6(c) and 6(d) show that when these simulations were carried out in the ideal case, with no line resistances or device variations, so that $g_{off}$ and $TMR$ are the same value for all devices, the network accuracy maximizes at the same value of $g_{norm}$ that minimizes the RMS deviation. Similar comparisons of simulated accuracies and RMS deviations as a function of $g_{norm}$ for different MTJ diameters with and without line resistances and device variations are provided in Figs. S5 and S6.

These simulations reveal that the line resistances amplify the effects of the device variations. We found that in simulations that neglect line resistances, to get results close to those in experiment, the device variations needed to be much larger than would be consistent with the distributions measured from single device measurements. Without line resistances, the voltage drop across every MTJ on a row was the same. However, when line resistances are included, the voltage drops across each device depend on the state (parallel or antiparallel) of each MTJ along that row. In the simulations that include the line resistances, the variations in the devices taken from other measurements lead to good agreement between the experiments on the arrays and the simulations of the arrays.

The simulated maximum achievable accuracies as a function of the MTJ diameter obtained over 30 different realizations of the MTJ device variations and 300 different weight matrix solutions are shown in Fig. 6(e), along with the experimental maximum achievable accuracies. In the absence of device-to-device variations and line resistances, the mean value of the maximum accuracies of the 300 different weight solutions obtained is about 99 %. The device conductances scale quadratically with the MTJ diameter, causing a corresponding increase in the relative variations in the two-port conductances in the presence of fixed line resistances. These increased variations, along with lower write accuracies, reduce the accuracies in both the experimental data and simulation results shown in Fig. 6(e). Similar variations of simulated maximum achievable accuracies as device parameters are scaled, starting from the nominal values corresponding to the MTJ array with 30 nm diameter, are provided in Fig. S8.

Using the simulations, we also calculate the distribution of $\xi_{norm}$, which is the ratio of $g_{norm}$ at minimum RMS deviation to the $g_{norm}$ at maximum accuracy as a function of MTJ diameter, and show it in Fig. 6(f). In the ideal case, $\xi_{norm}$ is unity for all MTJ sizes because the $g_{norm}$ that minimized RMS deviation is always equal to the $g_{norm}$ that maximized accuracy. In the full model, as the relative variations of conductance increase with MTJ diameter so does the disagreement between the $g_{norm}$ value that maximizes accuracy and the $g_{norm}$ value that minimizes RMS deviation, as seen by the increasing magnitude of $\xi_{norm}$ with MTJ diameter in both experimental and simulation data. By reproducing this disagreement in variational simulations, we posit that maximizing the network fidelity is not the same as maximizing the network accuracy for neural networks. This observation has important implications in embedded inference applications and suggests new techniques to compare a hardware recreation of a network to its software source are needed to improve the resilience of these systems.

## III. DISCUSSION

In this work, a 15 × 15 passive MTJ array was fabricated and programmed to analyze a hardware implementation for inference of a binary neural network trained to classify the Wine dataset [51]. To investigate the role played by hardware non-idealities, 300 unique weight matrix solutions were programmed into the array using a write-verify process and the accuracy was determined from the read conductance values. As expected, certain weight solutions perform better than others, but we find the accuracy values can be boosted significantly by optimizing the normalization conductance value. Surprisingly, the value of normalization conductance that minimizes the median of weight RMS deviation is not the same value that maximizes the median classification accuracy over all 300 weight solutions. These findings provide insight into the problem of embedded inference with MTJ-based hardware accelerators and are an integral step forward on the pathway toward large-scale integration of hardware devices with imperfections and variations. In reference [36], the MTJ network accuracies were about 1 % to 2 % below the baseline accuracy, however, as we show in this work, studying an *ensemble* of network solutions reveals a distribution of performance levels. Consequently, a broader exploration of the solution space compiled onto the crossbar and additional optimizations can likely lead to equivalent performance to the targeted baseline.

With this in mind, the work described here involves a small prototype array and a simple two-layer neural network applied to a very basic dataset. The results obtained, however, have important implications if scaling of this type of neural network is to be pursued. Any full-scale realization of a neural network using a passive MTJ array will necessarily include supporting complementary metal-oxide semiconductor (CMOS) circuitry. Co-design of the supporting circuits with the array is important since the properties of the former affect the overall accuracy, the overall power, and the requirements on the properties of the devices used in the array itself.

One important co-design constraint is the precision of the readout circuits. For single-bit precision in integrated binary neural network proposals [57], sense amplifiers are commonly used as thresholding elements, as used in commercial MTJ memory arrays. As neural networks are scaled to higher precisions, the readout circuits require a better signal-to-noise ratio, skewing the ratio of the power spent in the system toward the supporting circuits. Extending the sense amplifier to multibit precision can involve using analog to digital conversion techniques (such as successive approximation or flash). In this approach, the power dissipation scales as the square of the signal to noise ratio (SNR) [58]. In order to overcome this problem and improve output precision without significantly increasing the power of the amplifier, a recent work proposes using time-domain readout by measuring the $RC$ charging time of the output line, where $R$ is determined by the MTJs in the column based on input devices [36].

Once the co-design of the readout circuits is optimized with respect to the array, the following considerations determine the optimal array size. For a fixed current budget (and, hence, bandwidth) of these sense amplifiers, and a fixed $TMR$ of the MTJs, a successful VMM operation needs a high signal-to-noise ratio, which depends on the size of the array and the line resistances. For a fixed MTJ $TMR$, larger arrays require a proportional reduction in the line resistances. This ensures that, during array operation, most of the voltage is dropped across the MTJs. Once the physical limits of scaling line resistances are reached, the array size could be further increased by increasing the MTJ resistances (while keeping the $TMR$ fixed) until the bandwidth constraints on the sense amplifier are reached. The resulting array size is optimal. Scaling the VMM to larger sizes does not improve the performance of the system, since it runs into the bandwidth limitations of the sense amplifiers, while a smaller and faster VMM does not take full advantage of the available power budget.

Practically speaking, while the line resistances of the row and column lines used in this array are about 6 Ω per square, standard CMOS back-end-of the line processes using a dual damascene process are capable of producing interconnects with sheet resistances less than 1 Ω per square [59,60]. The resistance-area-product ($RA$) of the MTJs used in this study is about 20 $\Omega\cdot\mu m^2$. Higher MTJ resistances can be achieved, for example, by scaling down the MTJ diameter and by increasing the tunnel barrier thicknesses. While decreasing the diameter is limited by the thermal stability of the free layer magnetization, increasing the tunnel barrier thickness requires increased switching voltages [61]. By varying the MTJ stack composition and processes involved during fabrication, $RA$ values from a few to about 500 $\Omega\cdot\mu m^2$ capable of voltage switching [61,62], and $RA$ values of several $k\Omega\cdot\mu m^2$ to a few $M\Omega\cdot\mu m^2$ with field-assisted switching have been reported [63,64].

These considerations suggest that scaling of a passive MTJ array of the type investigated here is possible up to an optimal size, even when the necessary peripheral CMOS circuitry is included. Determination of the optimal size will involve detailed engineering design that considers all the necessary circuitry and the specifics of the semiconductor process to be used. In Ref. [36] a significantly larger array was implemented, and showed only minimal degradation in performance as compared to the ideal benefit. An important factor in this improvement is both the larger network size as well as the reduced line resistance. Nevertheless, it is likely that significant energy savings over conventional von Neumann-limited, software-based approaches will be realized by implementing large neural networks using this type of array.

## ACKNOWLEDGMENTS

We acknowledge Michael Grobis, Chris Petti, and Alexei Bogdanov from Western Digital for their contributions in the design of the MTJ arrays used in this study. This work was funded by The National Institute of Standards and Technology. Nitin Prasad is supported by Quantum Materials for Energy Efficient Neuromorphic Computing, an Energy Frontier Research Center funded by the U.S. DOE, Office of Science, Basic Energy Sciences, under Award DE-SC0019273. Advait Madhavan acknowledges support under the Cooperative Research Agreement Award No. 70NANB14H209, through the University of Maryland. This work benefited in part from materials and software developed under DARPA / ONR grant No. N00014-20-1-2031.

## APPENDIX A: MTJ ARRAY FABRICATION

The fabrication of a 15x15 MTJ device array started by defining the bottom wordlines, beginning with sputter depositing 5 nm of aluminum oxide followed by 200 nm of TaN (≈200 $\mu\Omega\cdot cm$) onto a thermally oxidized silicon substrate. The thin aluminum oxide layer acted as an etch stop during reactive ion etching (RIE) of the TaN so the thermal silicon oxide surface of the substrate was not attacked. Photolithography (i-line) was then used to pattern photoresist into alignment marks, metrology features used in subsequent process steps, and active array regions of 400 nm wide lines with a full pitch of 800 nm. This photoresist was used as a mask for RIE of the TaN, which had to be completely etched down to the aluminum oxide stop layer otherwise the wordlines could have been potentially shorted. Following wordline RIE, we sputter-deposited 200 nm of $SiO_2$ onto the wafer. Defining the wordlines concluded by performing chemical-mechanical polishing (CMP) to remove material from the wafer until left with exposed TaN wordlines approximately 120 nm thick, co-planar with the silicon oxide refill and quite smooth (RMS roughness ≈ 0.3 nm). Low RMS roughness for the surface on which the MRAM film is deposited is crucial for achieving the desired electrical and magnetic properties of the MTJs.

Prior to MRAM film deposition an *in situ* sputter etch removed any oxidized TaN at the surface of the wordlines, allowing for deposition directly onto low resistance TaN. The bottom-pinned MRAM stack was sputter deposited in a physical vapor deposition system, and then annealed in vacuum at 335 °C for 1 hour. A detailed description of the MRAM film stack including magnetic properties is given in Ref. [65].

After the deposition of the MRAM film, the hardmask for etching the film was deposited and patterned. TaN (40 nm) was sputter deposited followed by 50 nm of diamond-like carbon (DLC) and a final layer of 10 nm Cr. An aligned e-beam exposure then patterned an array of holes in high-resolution negative resist (HSQ) where the MRAM pillars were to be located. The various hardmask layers were then etched successively using a chlorine-based RIE process to transfer the patterns from HSQ into the Cr layer, then a $CO_2$ RIE process to etch the DLC without attacking the Cr, and finally a $CHF_3$/$CF_4$/Ar-based RIE process to etch the TaN without attacking either the Cr or DLC. DLC has excellent ion milling selectivity and provided most of the masking during ion milling of the MRAM stack. The primary purpose of the TaN layer was to act as a conductive cap since this material would be at the surface after subsequent steps described below.

After the hardmask was patterned, a multi-angle ion milling process was used to etch the MRAM stack down to the wordline layer. It was important to avoid incomplete milling as this would leave all the wordlines shorted together by the residual MRAM stack. Overmilling also needed to be avoided lest the mill penetrate too deeply into the wordlines. The final milling step was an oblique angle 200 V cleaning step to ensure redeposited metal was removed from the device sidewalls since it could possibly short the tunnel junction if present.

Following ion milling, the MRAM pillars were encapsulated with 5 nm of ion-beam-deposited aluminum oxide. $SiO_2$ (200 nm) was then sputter-deposited to fully encapsulate the MRAM bits. Using CMP we first planarized the wafer and then continued polishing until roughly midway into the TaN hardmask layer. This TaN served as a self-aligned via connecting the bitline directly to the MRAM devices.

The final steps in the process defined the vias, bitlines, and probe pads necessary for electrically connecting the pillars. Photolithography and RIE were first used to pattern vias in the $SiO_2$/alumina so the top electrodes would be able to electrically contact the wordlines beneath the MRAM bits at landing pads defined during wordline processing. Following the via etch, we sputter-deposited 5 nm Cr/≈1.2 nm Au/5 nm Cr onto the wafer, where the thickness of each metal was chosen such that the total resistance of the bitlines patterned from this film would match the resistance of the bottom TaN wordlines. Prior to depositing this metal film, an in-situ ion mill removed any oxide at the surface to ensure good electrical contact between the Cr and the TaN. Optical lithography was then used to pattern bitlines 400 nm wide on an 800 nm full pitch. The photoresist acted as a mask for the ion milling used to pattern the Cr/Au/Cr into the bitlines and was stripped using solvents following the etch step (see Fig. 1(a) for an image of the active region of one such array after completion of bitline processing). The process was completed with an optical lithography step accompanied by a Ta/Au deposition to pattern probe pads connected to the wordlines and bitlines used to make electrical measurements on devices.

## APPENDIX B: VERIFICATION OF CURRENT LINEAR SUPERPOSITION

In general, MTJs are non-linear elements with voltage-dependent resistances [66]. However, in the MTJ array considered in this work, the voltage drop seen across each MTJ is small enough that we can neglect these non-linearities and consider the MTJs to be linear elements. This allows us to analyze the full VMM element-by-element by applying a read voltage one row at a time. The resulting currents are then added up with appropriate scaling to obtain an approximation for the full VMM current output. In this Appendix, we describe experiments to test the linearity of the MTJ array. In these experiments, we compare the full VMM currents against those computed by summing the currents produced by applying read voltages one row at a time.

In the first set of measurements, we constructed 100 input vectors $\vec{x}_{inputs}$ of size 15 with elements $x_i$ being 0 or 1 chosen randomly with equal probability. A voltage was then applied simultaneously on all rows where $x_i = 1$ while rows with $x_i = 0$ were grounded. The applied voltage was varied from 0.1 V to 0.5 V in steps of 0.1 V. Current $I_{j,p}$ on column $j$ was measured at each applied voltage for all 100 $\vec{x}_{inputs}$. We refer to these measurements as "parallel" measurements, indicated by the subscript $p$. The full VMM of each of the 100 $\vec{x}_{inputs}$ can be obtained from these $I_{j,p}$.

In a second set of measurements, we perform a current read on each individual device $I_{ij,s}$ by applying a read voltage (0.2 V) on the $i$th row and measuring the current on the $j$th column while all other connections are grounded. We refer to these measurements as "serial" measurements, indicated by the subscript $s$. An approximation of the full VMM was then obtained for comparison.

Figure 7 plots the relative RMS deviation between the current vector for devices measured in parallel and serially, where the individual currents were calculated from the device conductance as measured at 0.2 V. The relative RMS deviation for Fig. 7 is calculated as:

$$\text{Relative RMS Deviation} = \frac{\sqrt{\sum_j \left(I_{j,p} - \sum_i x_i \, g_{ij,s\,@\,0.2\,V} \, V_{apply}\right)^2}}{\sqrt{\sum_j \left(\sum_i x_i \, g_{ij,@\,0.2\,V} \, V_{apply}\right)^2}}, \quad \text{(B1)}$$

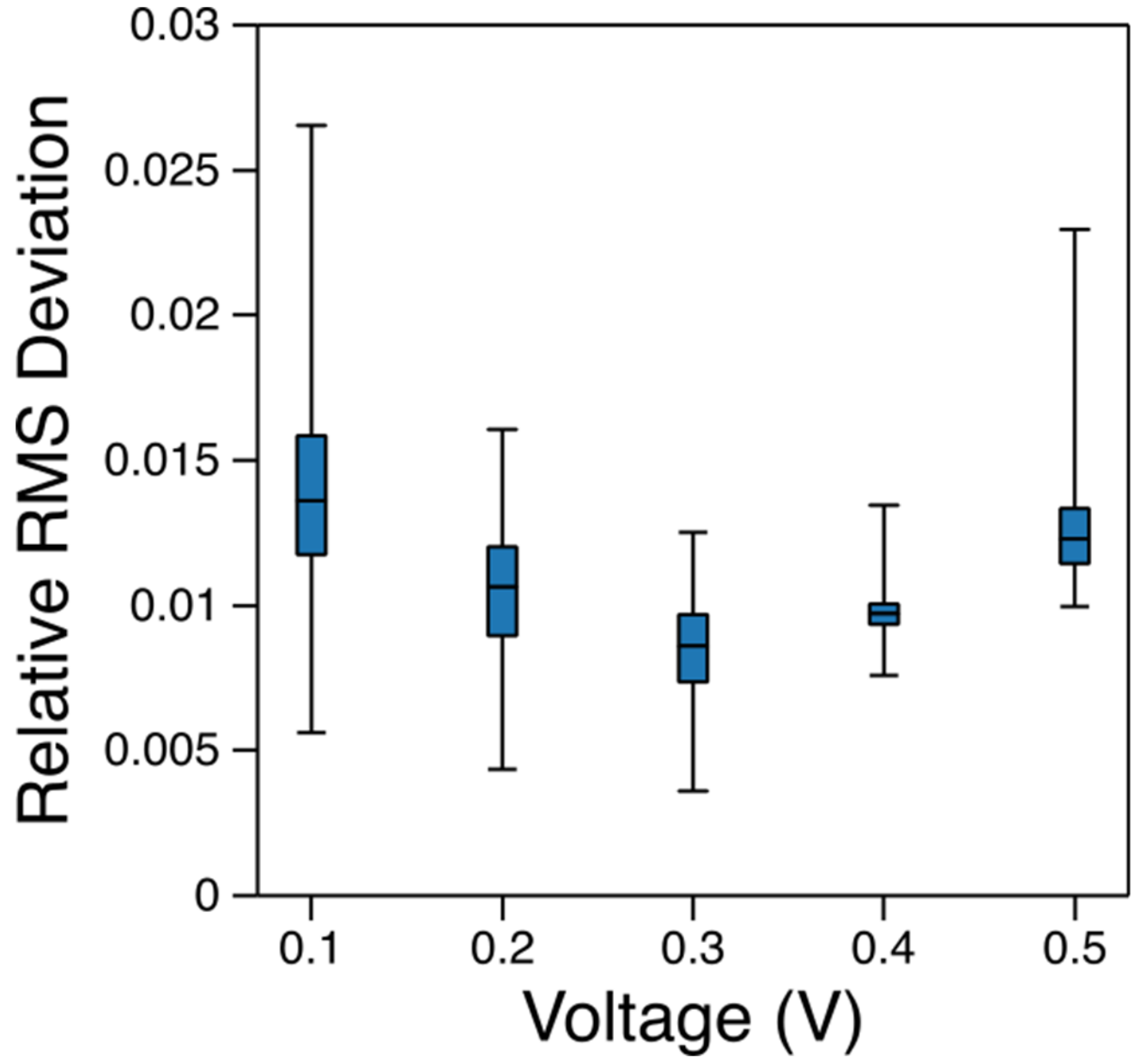

FIG. 7. Box-whisker plots showing the distribution over 100 randomizations of connected rows of the relative RMS deviation between the current vector output of the 15 columns based on the conductance of individual devices read at 0.2 V as a function of applied voltage. Whiskers indicate maximum and minimum values, whereas box edges represent 25 % and 75 % quartiles and the middle line is the median (50 % quartile).

where $g_{ij,s\,@\,0.2\,V}$ is the conductance of an individual device measured at 0.2 V. The deviation is less than 3 % for applied voltages up to 0.5 V and more typically about 1 %. In addition, the standard deviation in the measurements of the current of each individual device is ≈10 nA, giving a measurement uncertainty of ≈1 %, which accounts for nearly half the measurement error in Fig. 7. This shows that even if there is some error on the current of individual columns, the error in the output vector is below 2 %, and based on Fig. 1(d), this does not change the accuracy significantly. These results demonstrate that the port-to-port resistance of individual devices in the array is sufficiently linear with applied voltage and the assumption that linear superposition applies to this system is valid. It also shows the devices are sufficiently linear with respect to the conductance values at 0.2 V and the simulation described in the main text provides valid conclusions. Each of these measurements was performed on all sizes of MTJ arrays used in the paper and the findings are identical (see full set of figures in Fig. S7).

# Supplementary Materials for

## Implementation of a Binary Neural Network on a Passive Array of Magnetic Tunnel Junctions

Jonathan M. Goodwill, Nitin Prasad, Brian D. Hoskins, Matthew W. Daniels, Advait Madhavan, Lei Wan, Tiffany S. Santos, Michael Tran, Jordan A. Katine, Patrick M. Braganca, Mark D. Stiles, Jabez J. McClelland

**This PDF file includes:**

Figs. S1 to S8

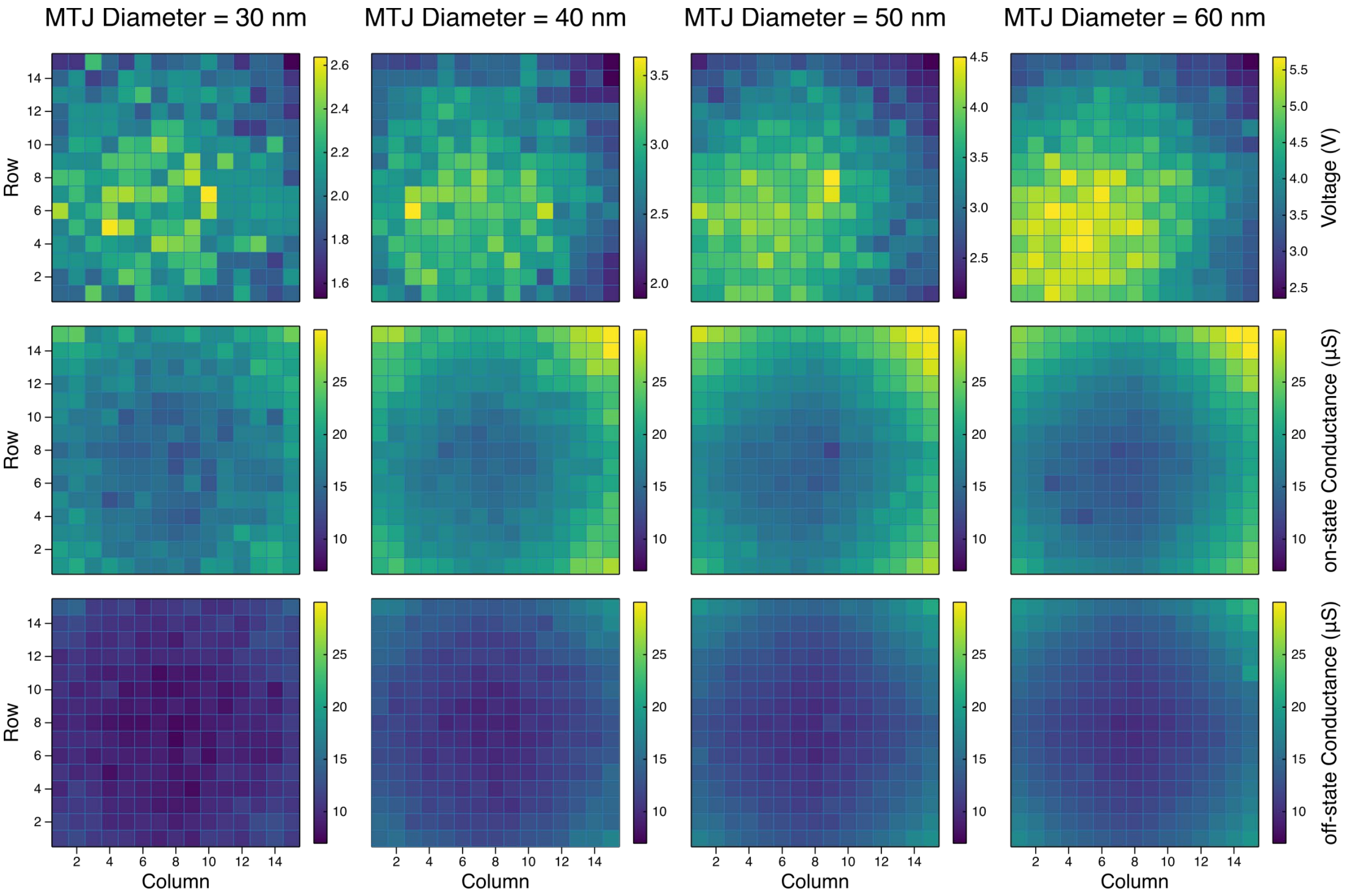

Fig. S1. Map of average switching voltage, on-state conductance, and off-state conductance over all 300 tested weight solutions as a function of MTJ device size. Voltage magnitude increases with increasing MTJ size. On- and off-state conductance states are plotted with the same colorbar range.

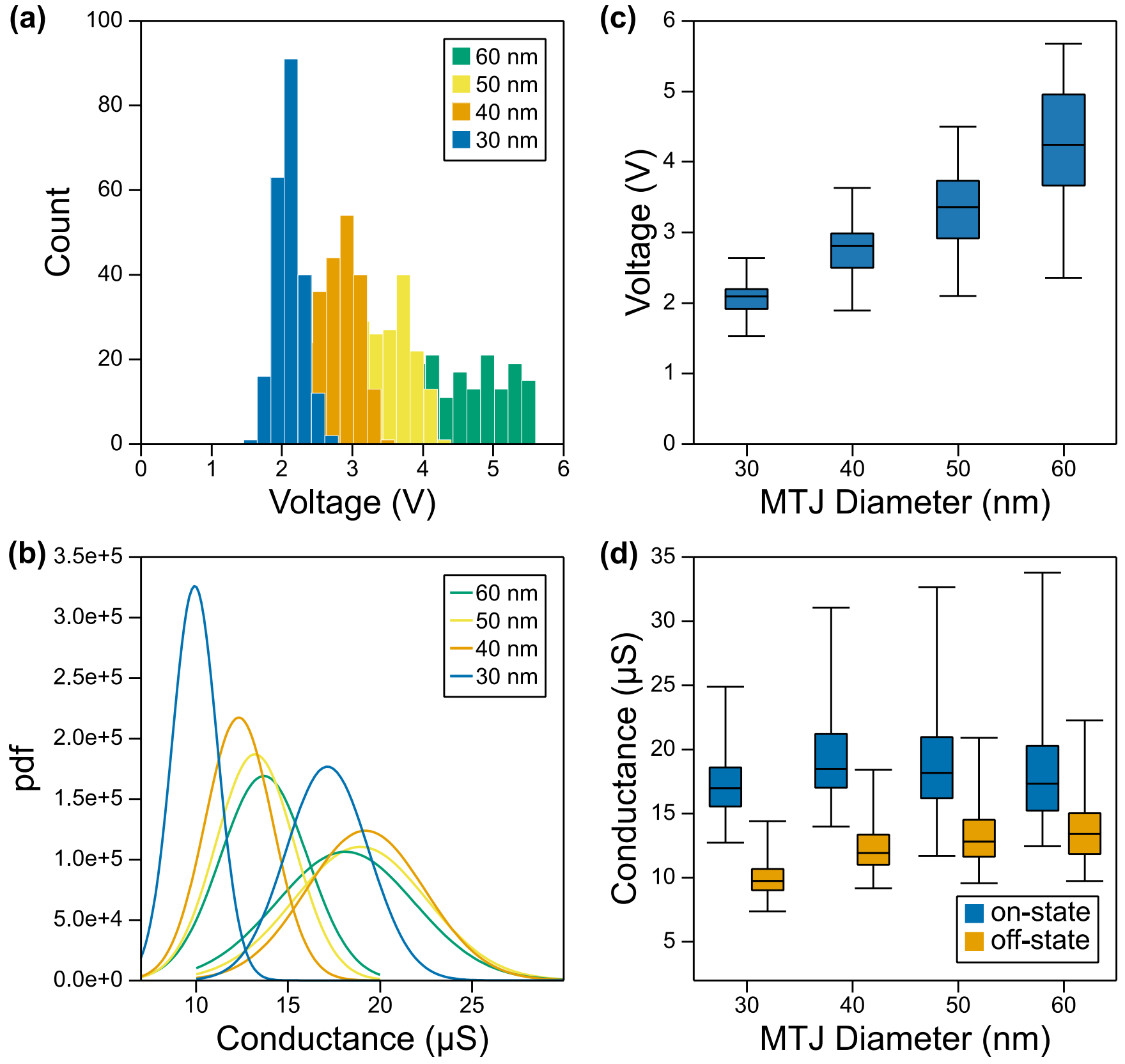

Fig. S2. Voltage and on/off-states conductance characteristics for different MTJ sizes. (a) Histogram of MTJ switching voltage as a function of MTJ diameter. (b) Probability density function for on- and off-state conductances of each MTJ size. (c) Box-whisker plot showing the distribution of switching voltage as a function of MTJ size. (d) Box-whisker plot of on- and off-states for each MTJ size. In (c) and (d), whiskers indicate maximum and minimum values, whereas box edges represent 25 % and 75 % quartiles and the middle line is the median (50 % quartile).

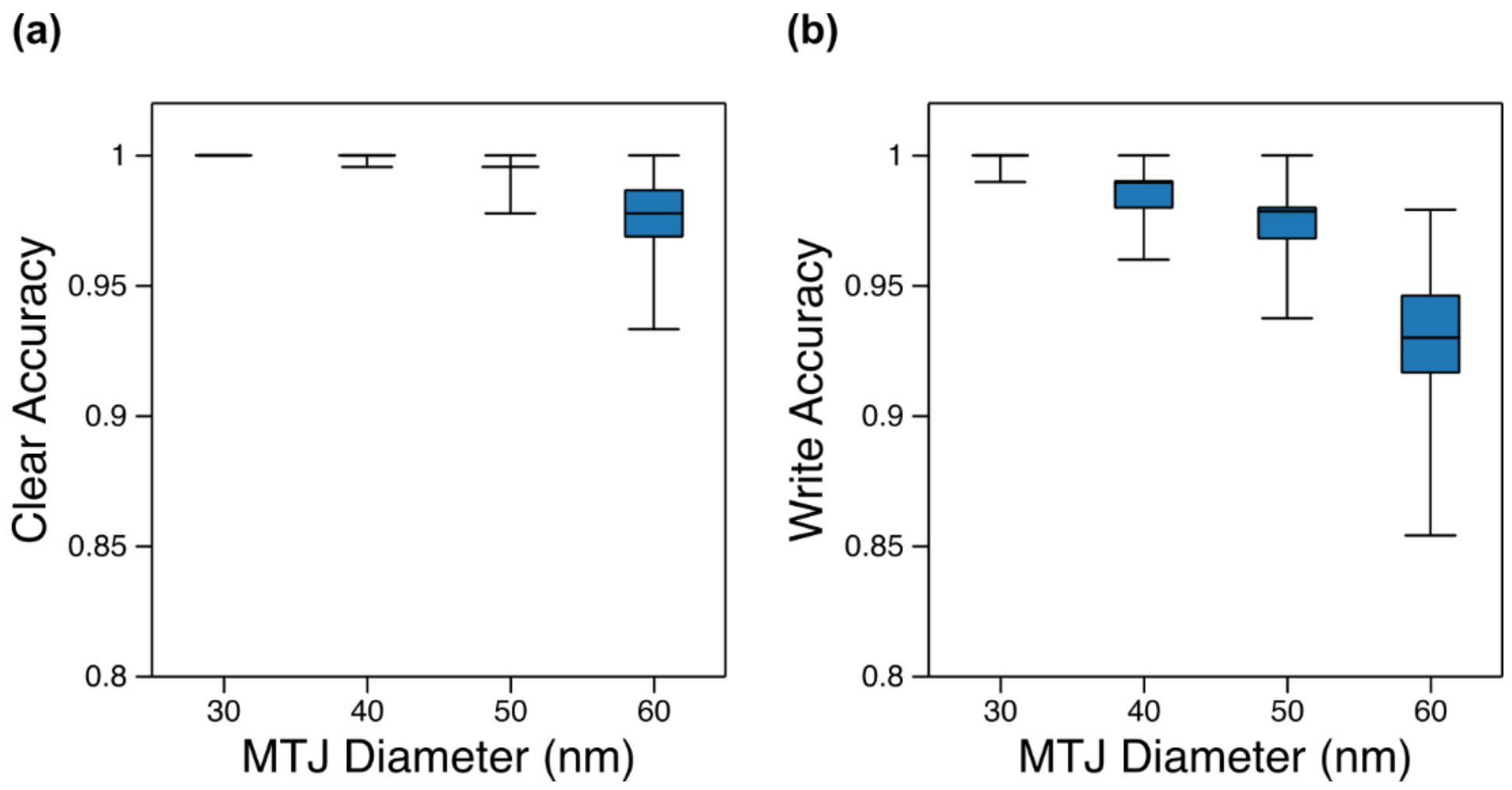

Fig. S3. Box-whisker plots of distributions of clear and write accuracy for different MTJ sizes. (a) Successful clear accuracy as a function of MTJ size. Clear accuracy was defined as the number of successful clears out of all 225 devices. (b) Successful write accuracy as a function of MTJ size. Write accuracy was defined as the number of successful writes out of attempted writes. After the "clear operation," only devices with a target on-state were attempted to be written. In each figure, whiskers indicate maximum and minimum values, whereas box edges represent 25 % and 75 % quartiles and the middle line is the median (50 % quartile).

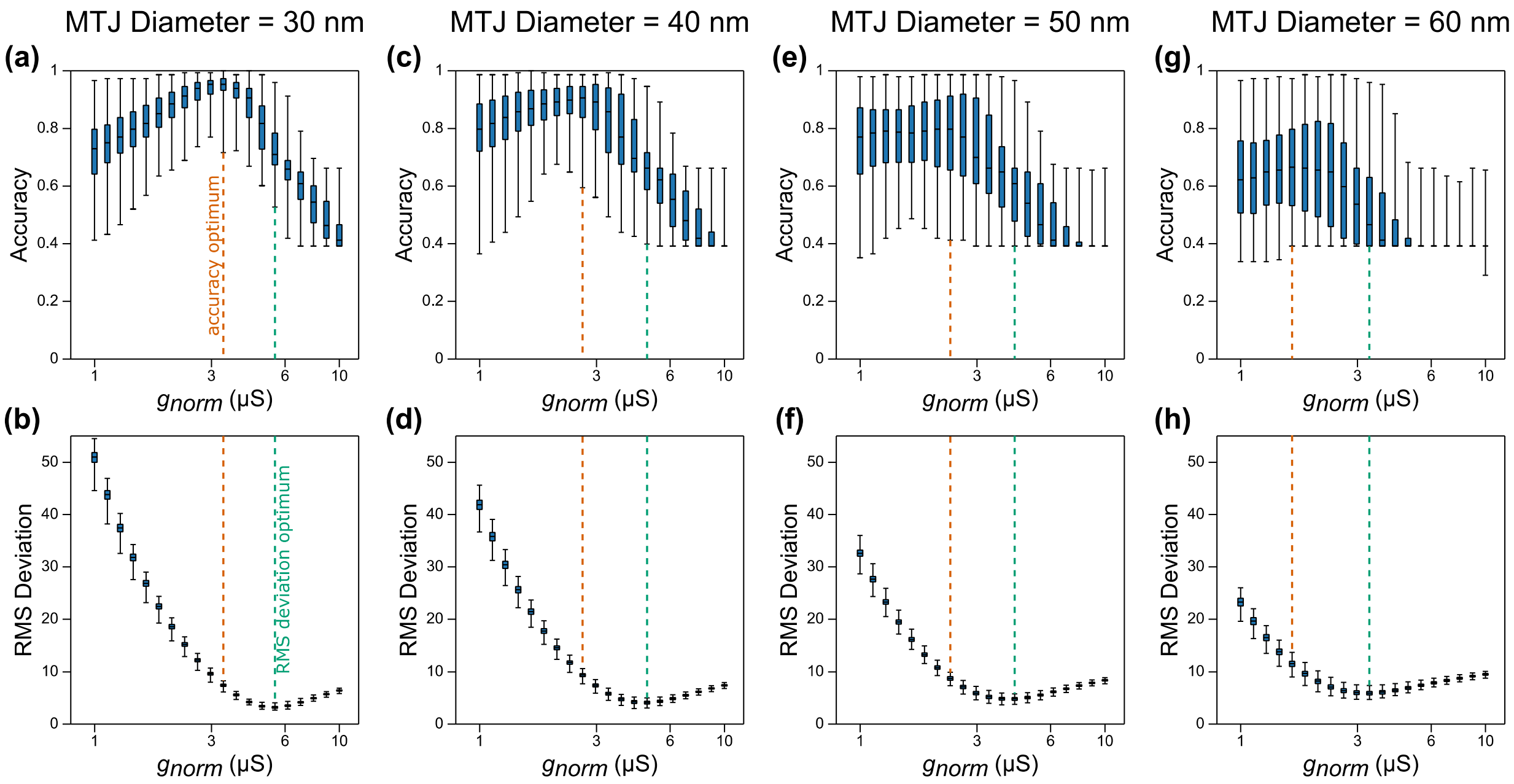

Fig. S4. Box-whisker plots of experimental (a,c,e,g) inference accuracy and (b,d,f,h) RMS deviation as a function of $g_{norm}$ for all MTJ device sizes. In each figure, whiskers indicate maximum and minimum values, whereas box edges represent 25 % and 75 % quartiles and the middle line is the median (50 % quartile).

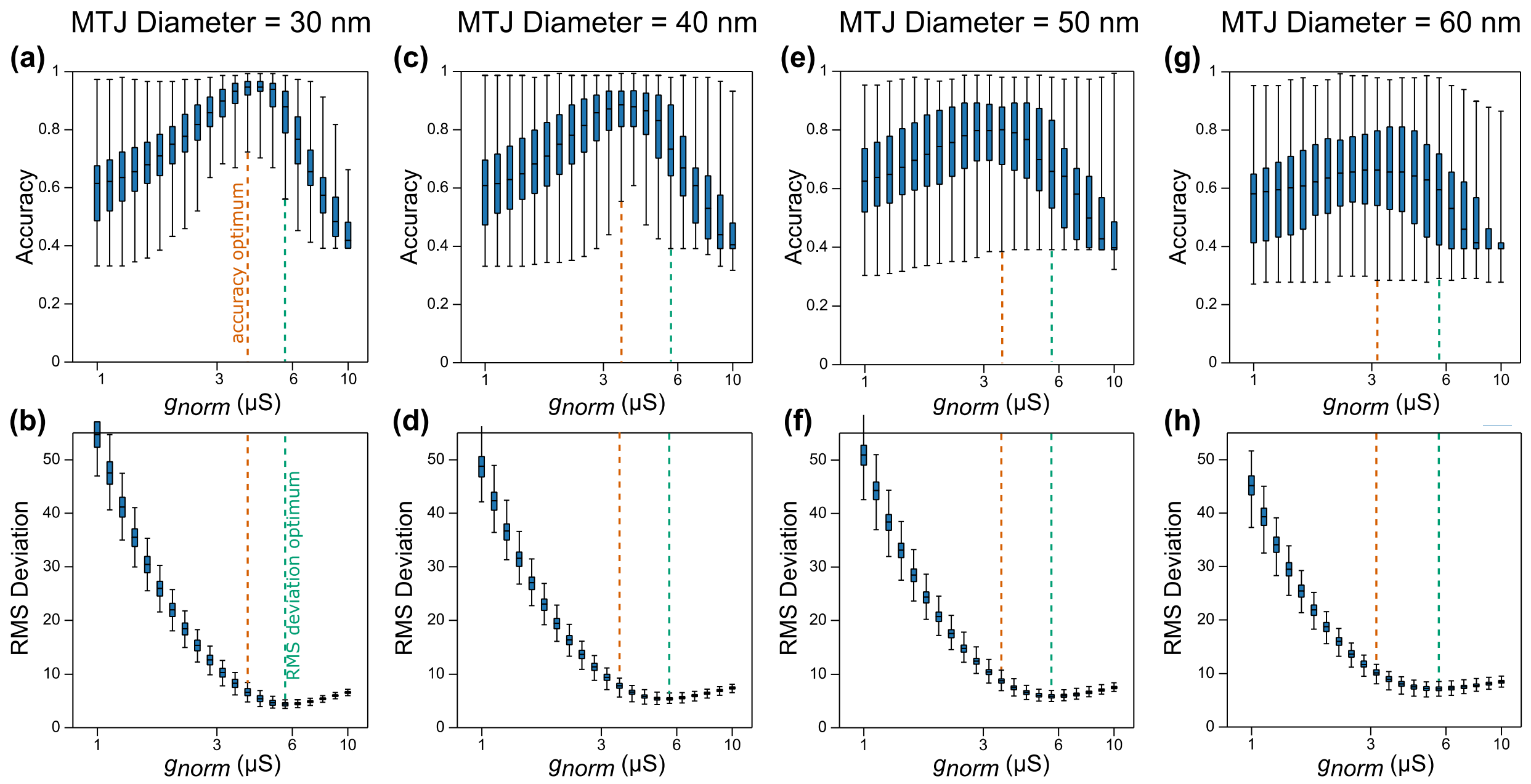

Fig. S5. Box-whisker plots of the distribution of results over the 300 unique weight solutions of (a,c,e,g) inference accuracies and (b,d,f,h) RMS deviations as a function of $g_{norm}$ in the presence of device conductance variations and line resistances. These results correspond to a representative realization of normally distributed device conductance variations for each MTJ diameter. In each figure, whiskers indicate maximum and minimum values, whereas box edges represent 25 % and 75 % quartiles and the middle line is the median (50 % quartile).

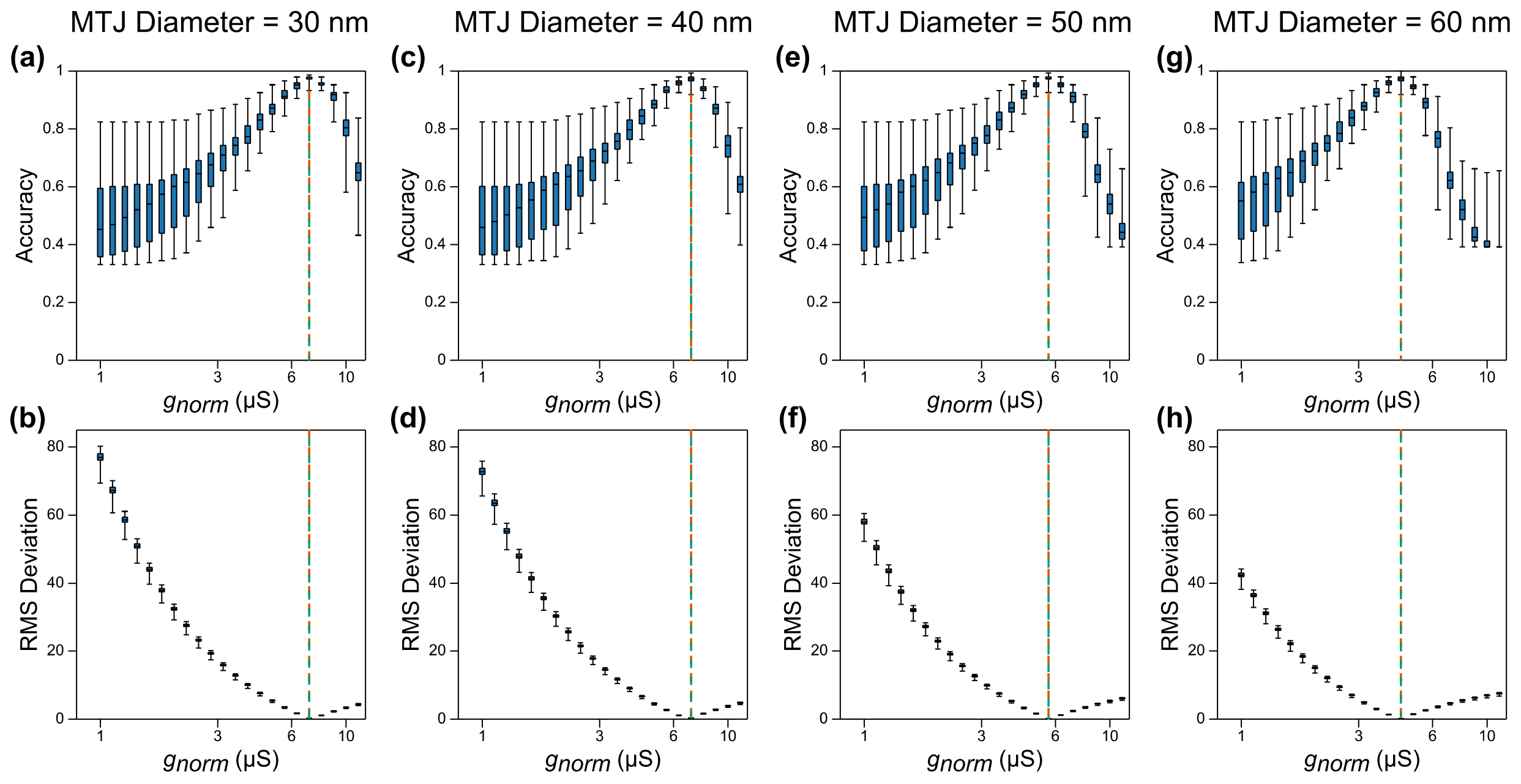

Fig. S6. Box-whisker plots of simulated (a,c,e,g) inference accuracies and (b,d,f,h) RMS deviations as a function of $g_{norm}$ assuming the ideal situation where all devices have the same off-states and $TMR$ equal to average values for each MTJ array. In each figure, whiskers indicate maximum and minimum values, whereas box edges represent 25 % and 75 % quartiles and the middle line is the median (50 % quartile).

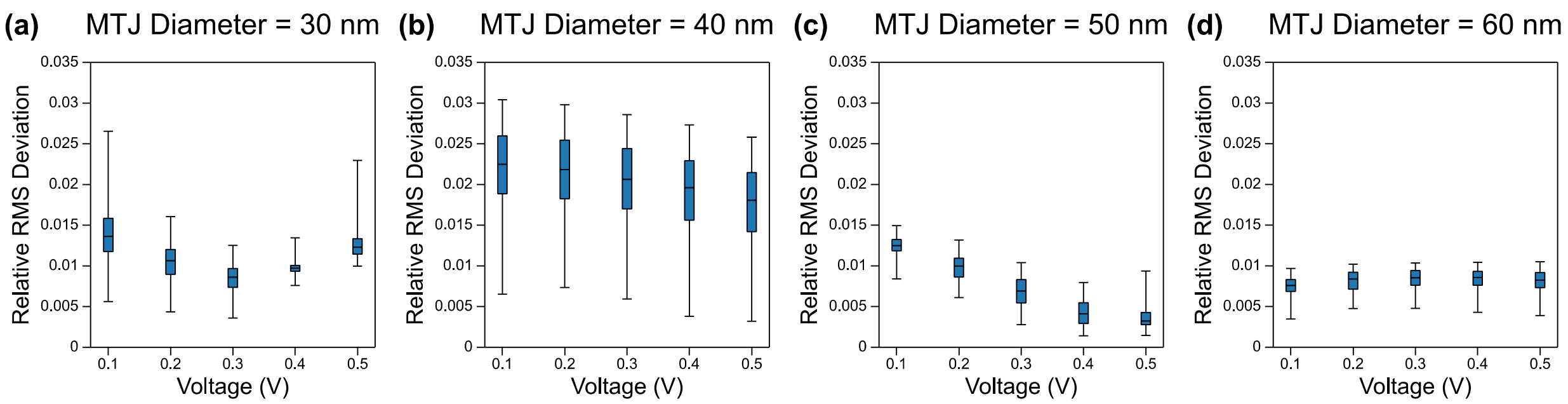

Fig. S7. Box-whisker plots of the relative RMS deviation of the column current vector output of different MTJ sizes using individually measured device conductances at 0.2 V versus the devices measured in parallel for all 100 randomized row connections as a function of applied voltage. In each figure, whiskers indicate maximum and minimum values, whereas box edges represent 25 % and 75 % quartiles and the middle line is the median (50 % quartile).

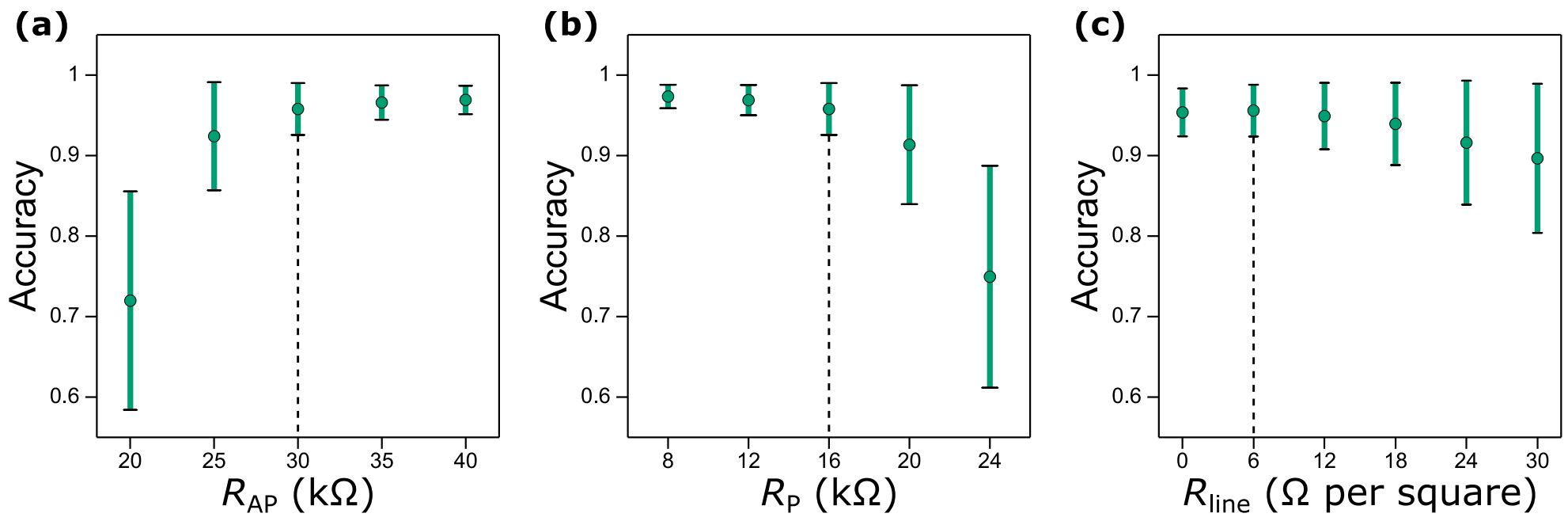

Fig. S8. Simulated accuracy distributions as the mean values of (a) MTJ resistance in the antiparallel configuration, $R_{AP}$, (b) in the parallel configuration, $R_P$ and (c) the sheet resistance $R_{line}$ are individually varied starting from a model representing the 30 nm MTJ array. The dashed line corresponds to the nominal fit value used to model the 30 nm MTJ array. The solid circles correspond to mean accuracies while the whiskers represent one standard deviation from the mean.